\documentclass[aps,prd,onecolumn,groupedaddress,amssymb,showpacs,nofootinbib,epsfig,pdffig]{revtex4}
\usepackage{graphicx}
\usepackage{bm}
\usepackage{dcolumn}
\usepackage{amsmath}
\usepackage{amssymb}
\usepackage{epsfig}
\usepackage{color}

\def \lleq {\lower0.9ex\hbox{ $\buildrel < \over \sim$} ~}
\def \ggeq {\lower0.9ex\hbox{ $\buildrel > \over \sim$} ~}

\def \beq  {\begin{equation}}
\def \eeq  {\end{equation}}
\def \ber  {\begin{eqnarray}}
\def \eer  {\end{eqnarray}}

\newcommand{\be}{\begin{equation}}
\newcommand{\ee}{\end{equation}}
\newcommand{\ba}{\begin{eqnarray}}
\newcommand{\ea}{\end{eqnarray}}
\newcommand{\bea}{\begin{eqnarray*}}
\newcommand{\eea}{\end{eqnarray*}}

\begin{document}

\title{Late time evolution of a nonminimally coupled scalar field system }

\author{  M. Shahalam$^{1,2}$, R. Myrzakulov$^3$, Maxim Yu. Khlopov$^{4,5,6}$
}
\affiliation{$^{1}$Institute for Theoretical Physics $\&$ Cosmology, Zhejiang University of Technology, Hangzhou- 310023, China}
\affiliation{$^{2}$Centre for Theoretical Physics, Jamia Millia Islamia,
New Delhi-110025, India}
\affiliation{$^{3}$Eurasian International Center for Theoretical Physics, Department of General and Theoretical Physics, Eurasian National
University, Astana, Kazakhstan}

\affiliation{$^{4}$National Research Nuclear University “MEPHI” (Moscow Engineering Physics Institute), 115409 Moscow, Russia}
\affiliation{$^{5}$APC Laboratory, 10 rue Alice Domon et Léonie Duquet, 75205 Paris Cedex 13, France}
\affiliation{$^{6}$Institute of Physics, Southern Federal University,Rostov on Don, Russia}

\begin{abstract}
We revisit the dynamics of a nonminimally coupled scalar field model in case of $F(\phi)R$ coupling with $F(\phi)= 1-\xi\phi^2 $, the potentials $V(\phi) = V_0 (1+ \phi^p)^2$ and 
 $V(\phi)= V_0 e^{\lambda \phi^2}$. We use an autonomous system to bring out new asymptotic regimes, and find stable de-Sitter solution. Under the chosen functional form of $F(\phi)$  and steep exponential potentials, a true de-Sitter solution is trivially satisfied for which the equation of state $w_{\phi}\simeq -1$, the effective gravitational constant $G_{eff}$ and field $\phi$ are constant that has been missed in the power law case and our previous study.
\end{abstract}
\maketitle

\section{Introduction}
\label{sec:intro}
The scalar fields play a vital role in cosmology. They are used in quintessence, inflation, and in the dynamics of loop quantum cosmology (LQC) etc. \cite{review2,alamLQC}. In the framework of LQC, the initial evolution of scalar field is divided into two different classes, one is dominated by kinetic energy and other is not. For kinetic energy dominated case (except a small subset), the evolution of universe before reheating can be categorized into three distinct phases: bouncing, transition and slow-roll inflation. This universal characteristic is independent for a wide range of initial conditions and specific potentials of scalar field, as long as it is dominated by kinetic energy \cite{alamLQC}. In context of modified LQC, the analytical solution with various approximation are obtained for different physical variables and compared by the numerical ones in case of kinetic energy dominated initial conditions of inflaton field. The universal feature of dynamics, shared with LQC, are established and properties of three different phases: bouncing, transition and slow-roll inflation are studied \cite{wangLQC}.

The energy density of the minimally coupled scalar field to gravity imitate the effective cosmological behavior. We make the extension in quintessence by including the nonminimally coupled (NMC) scalar field to gravity. In the literature, it is known as the scalar-tensor theory that has been studied for decades, and appeared in Brans-Dicke theory to match the Mach's principle with general relativity \cite{BD}. In this theory, the Newtonian gravitational constant is not a constant but it is the function of scalar field that appears into the action in a particular form with the curvature term as $\phi^2 R$. The NMC scalar field models due to interesting features are of great interest to dark energy models \cite{PR1,PR2,review1,vpaddy,review3,review3C,review3d,review4}, and have been widely studied in \cite{a1,a2,a3,a4,a5,a6,a7,a8,a9,a10,a11,a12,a13,a14,a15,a16,sunny}. A familiar model of a NMC scalar field system is given by $F(\phi) R$ coupling with $F(\phi)=1- \xi \phi^2$. Many authors have applied the NMC scalar field model in the context of late time cosmology to address the dark energy problem as it avoids the coincidence problem, allows the phantom crossing in some cases, and may provide the cosmological scaling behavior. Phantom scaling solutions are generic characteristics of a NMC scalar field model having  $F(\phi)=1- \xi \phi^2$ \cite{Polarski}.

Note that the methods of dynamical system theory are extensively used in the literature to obtain a dynamical picture for various cosmological models. Using these methods many asymptotic solutions are obtained and their stability has been scrutinized with a simple programmed algorithm. Conclusively, a coalition of the phase portrait and the stability is a standard way to obtain the viable cosmological behaviors.

In this paper, we study the dynamics of a NMC scalar field model having a particular form of $F(\phi)$, power law and steep exponential potentials, and investigate the stationary points and their stability. We restrict ourselves to the functional form of  $F(\phi)$ as $F(\phi)=1- \xi B(\phi)$ with $B(\phi) \propto \phi^2$ and two different potentials such as power law and steep exponential. We shall choose same autonomous system as has been used in our previous paper for $B(\phi) \propto \phi^N$ $(N \geq 2)$ and power law potential $V(\phi) \propto \phi^n$ \cite{alam2012}. In the present study, we shall obtain new asymptotic regimes and generic features of the underlying dynamics that have been missed in our earlier paper. In Ref. \cite{alam2012}, we did not find a stable de-Sitter solution as $G_{eff}$ is negative in case of $B(\phi) \propto \phi^N$ $(N \geq 2)$ and $V(\phi) \propto \phi^n$. However, in the current study, we find a true de-Sitter solution as $G_{eff}$ and $\phi$ are constant for $B(\phi) \propto \phi^2$ and   $V(\phi) = V_0 e^{\lambda \phi^2}$ that trivially satisfied the de-Sitter conditions. However, in case of $V(\phi) = V_0 (1+ \phi^p)^2$, the stationary point is not stable in the usual sense.

The paper is organized as follows. In Section \ref{sec:EOM}, we discuss the background equations for a NMC scalar field model, and construct the autonomous system that is useful for phase space analysis. In Section \ref{sec:phase}, we obtain stationary points, stability and draw phase space trajectories for the model under consideration. We summarize our results in Section \ref{sec:conc}.

\section{Equations of motion}
\label{sec:EOM}
We consider the following action for a nonminimally coupled scalar field model \cite{alam2012}
 \be
 \label{eq:Lagrangian}
S=\frac{1}{2}\int{\sqrt{-g}d^4x\Big{[} m_{Pl}^2
R-(g^{\mu\nu}\phi_{\mu}\phi_{\nu}+ \xi R
B(\phi)+2V(\phi))\Big{]}}+S_M, \ee
where $m_{Pl}^2=({8\pi
G})^{-1}=({\kappa})^{-1}$,  the coupling constant $\xi$ is a dimensionless parameter  and $S_M$ denotes the matter action.

The equations of  motion in a spatially flat Friedmann-Leimetre-Robertson-Walker background are obtained by varying the action (\ref{eq:Lagrangian}), and given by
\be \label{eq:Friedphi} H^2=\frac{\kappa}{3}\left(\frac{1}{2}{\dot
{\phi}}^{2}+V(\phi)+3\xi(H \dot {\phi} B'(\phi)+H^{2}B(\phi))+\rho
\right), \ee
\begin{eqnarray}
\label{eq:Friedphi2} R=\kappa \left(-{\dot {\phi}}^{2} +4
V(\phi)+3\xi(3 H \dot{\phi}  B'(\phi)+\frac{R}{3}  B(\phi)  + {\dot
{\phi}}^{2}B''(\phi)+\ddot {\phi} B'(\phi))+\rho(1-3\omega) \right),
\end{eqnarray}
\begin{eqnarray}
\label{eq:KGphi}
&&\ddot {\phi}+3 H \dot {\phi}+\frac{1}{2}\xi R B'(\phi)+V'(\phi)=0.
\end{eqnarray}
where $p$ and $\rho$  are the pressure and energy density of the matter with $p=\omega \rho$.

From the standard form of equations
\begin{eqnarray*}
R_{ij}-\frac{1}{2}R g_{ij}&=&8\pi G_{eff}(T_{ij,\phi}+T_{ij,m})=\kappa
T^{eff}_{ij} ,
\end{eqnarray*}
The expression of the effective gravitational constant is given by \cite{alam2012}
\begin{equation}
G_{eff}=\frac{\kappa}{8\pi(1-\kappa \xi B(\phi))},
\end{equation}
We also define Ricci Scalar as $R=6(2H^2+ \dot {H})$. For the sake of simplicity,
we choose $\kappa=6$ \cite{alam2012}.

Dividing equations (\ref{eq:Friedphi}), (\ref{eq:Friedphi2}), (\ref{eq:KGphi})
by $H^2(1-6\xi B(\phi))$ and  multiplying equation (\ref{eq:KGphi}) by $\xi
B'(\phi)$, we have
\begin{eqnarray}
\label{eq:newFried1} 1=\frac{{\dot {\phi}}^{2} }{H^2(1-6\xi
B(\phi))}+\frac{2 V(\phi)}{H^2(1-6\xi B(\phi))} +\frac{6\xi \dot
{\phi} B'(\phi)}{H(1-6\xi B(\phi))}+\frac{2\rho}{H^2(1-6\xi
B(\phi))},
\end{eqnarray}
\begin{eqnarray}
\label{eq:newFried2} \frac{R}{H^2}&=&-\frac{6 {\dot {\phi}}
^{2}}{H^2(1-6 \xi B(\phi))}+\frac{24 V(\phi)}{H^2(1-6 \xi B(\phi))}
+\frac{54 \xi \dot {\phi} B'(\phi)}{H(1-6\xi B(\phi))}+\frac{18 \xi
{\dot{\phi}} ^2 B''(\phi)}{H^2(1-6 \xi B(\phi))}\nonumber\\
&+&\frac{18 \xi \ddot {\phi}  B'(\phi)}{H^2(1-6 \xi B(\phi))}
 +\frac{6\rho(1-3\omega)}{H^2(1-6 \xi B(\phi))},
\end{eqnarray}
\begin{eqnarray}
\label{eq:newKG} 0=\frac{\xi \ddot {\phi} B'(\phi)}{H^2(1-6 \xi
B(\phi))}+\frac{3 \xi \dot {\phi} B'(\phi)}{H(1-6 \xi B(\phi))}
+\frac{R}{H^2}\frac{\xi^{2}{B'}^2(\phi)}{2(1-6 \xi B(\phi))}
+\frac{V'(\phi)\xi B'(\phi)}{H^2(1-6 \xi B(\phi))}.
\end{eqnarray}
To cast above equations in an autonomous system, we choose the following dimensionless
parameters,
\begin{eqnarray}
\label{eq:omega} x &=& \frac{{\dot {\phi}}^{2}}{H^2(1-6 \xi
B(\phi))}, ~~y = \frac{2 V(\phi)}{H^2(1-6 \xi B(\phi))},~~
z = \frac{6\xi \dot {\phi} B'(\phi)}{H(1-6 \xi B(\phi))},\nonumber\\
\Omega &=& \frac{2 \rho}{H^2(1-6 \xi B(\phi))},~~ A=\frac{B'(\phi)\phi}{(1-6 \xi B(\phi))},~~ b=\frac{B''(\phi)\phi}{B'(\phi)},~~ c=\frac{V'(\phi)\phi}{V(\phi)},
\end{eqnarray}
where $'$ designates the derivative with respect to $\phi$.
Hence, the equations of motion (\ref{eq:newFried1}), (\ref{eq:newFried2}) and
(\ref{eq:newKG}) can be written as
\begin{eqnarray}
\label{eq:autonomous4}
\frac{dx}{d\ln a} &=& x' = 12X\frac{x}{z}-2x(\frac{Y}{6}-2)+xz,\nonumber\\
\frac{dy}{d\ln a} &=& y' = \frac{yz}{6\xi}\frac{c}{A}-2y(\frac{Y}{6}-2)+yz,\nonumber\\
\frac{dz}{d\ln a} &=& z' = 6X+\frac{z^2}{6\xi}\frac{b}{A}-z(\frac{Y}{6}-2)+z^2,\nonumber\\
\frac{dA}{d\ln a} &=& A' = \frac{z}{6\xi}(b+1)+Az,\nonumber\\
\frac{d\Omega}{d\ln a} &=& {\Omega}^{'} = \Omega(-3-3\omega-2(\frac{Y}{6}-2)+z),
\end{eqnarray}
The higher order derivative terms  having $\dot H$  and
$\ddot {\phi}$ in the autonomous system are
\begin{eqnarray}
\label{eq:xyr} X\equiv\frac{\xi \ddot {\phi} B'(\phi)}{H^2(1-6 \xi
B(\phi))},~~~ Y\equiv\frac{R}{H^2},
\end{eqnarray}
The expressions of $\Omega$, $X$ and $Y$ can also be expressed in terms of $x,y,z$ and
are given by
\begin{eqnarray}
\label{eq:capxy}
\Omega&=&1-x-y-z,\nonumber\\
X(x,y,z)&=&-\frac{z}{2}-\frac{z^2}{18(4x+z^2)}\left(-6x+12y+\frac{z^2
b}{2\xi A}+\frac{yc}{\xi A}\right.
 + 3 (1-x-y-z)(1-3\omega) \Big{)},\nonumber\\
Y(x,y,z)&=&\frac{4x}{4x+z^2}\left(-6x+12y+\frac{z^2}{4\xi A}
\left(2b-\frac{yc}{x}\right)\right. + 3 (1-x-y-z)(1-3\omega)\Big{)}.
\end{eqnarray}

The energy density, pressure and the equation of state for a NMC scalar field model are defined as
\begin{eqnarray}
\label{eq:rhophi}
\rho_{\phi}&=&\frac{1}{2}{\dot {\phi}}^{2}+V(\phi)+3\xi( H \dot {\phi} B'(\phi)+H^{2} B(\phi)),\\
p_{\phi}&=&\frac{1}{2}{\dot {\phi}}^{2}-V(\phi)-\xi\left(2 H \dot
{\phi} B'(\phi)+ \dot {\phi}^{2} B''(\phi)+\ddot {\phi}
B'(\phi)+(2\dot{H}+3H^{2}) B(\phi)\right).
\end{eqnarray}
\be
w_{\phi} = \frac{p_{\phi}}{\rho_{\phi}}= \frac{x-y-z-4 \xi x-2 X-2 \xi A (Y/6-1/2)}{x+y+z+3 \xi A}
 \ee

To obtain fixed points, we shall use autonomous system
(\ref{eq:autonomous4}), and would be interested in stable solutions that can give rise the late time cosmic acceleration. In what follows, we shall consider particular functional forms for the functions $B(\phi)$ and $V(\phi)$.

\section{Phase Space Analysis: Stationary points and their stability}
\label{sec:phase}
\subsubsection{Model 1: $B(\phi) \propto {\phi}^2$, $V(\phi) = V_0 (1+ \phi^p)^2$}
\label{sec:model1}
\begin{figure}[tbp]
\begin{center}
\begin{tabular}{ccc}
{\includegraphics[width=2.5in,height=2.5in,angle=0]{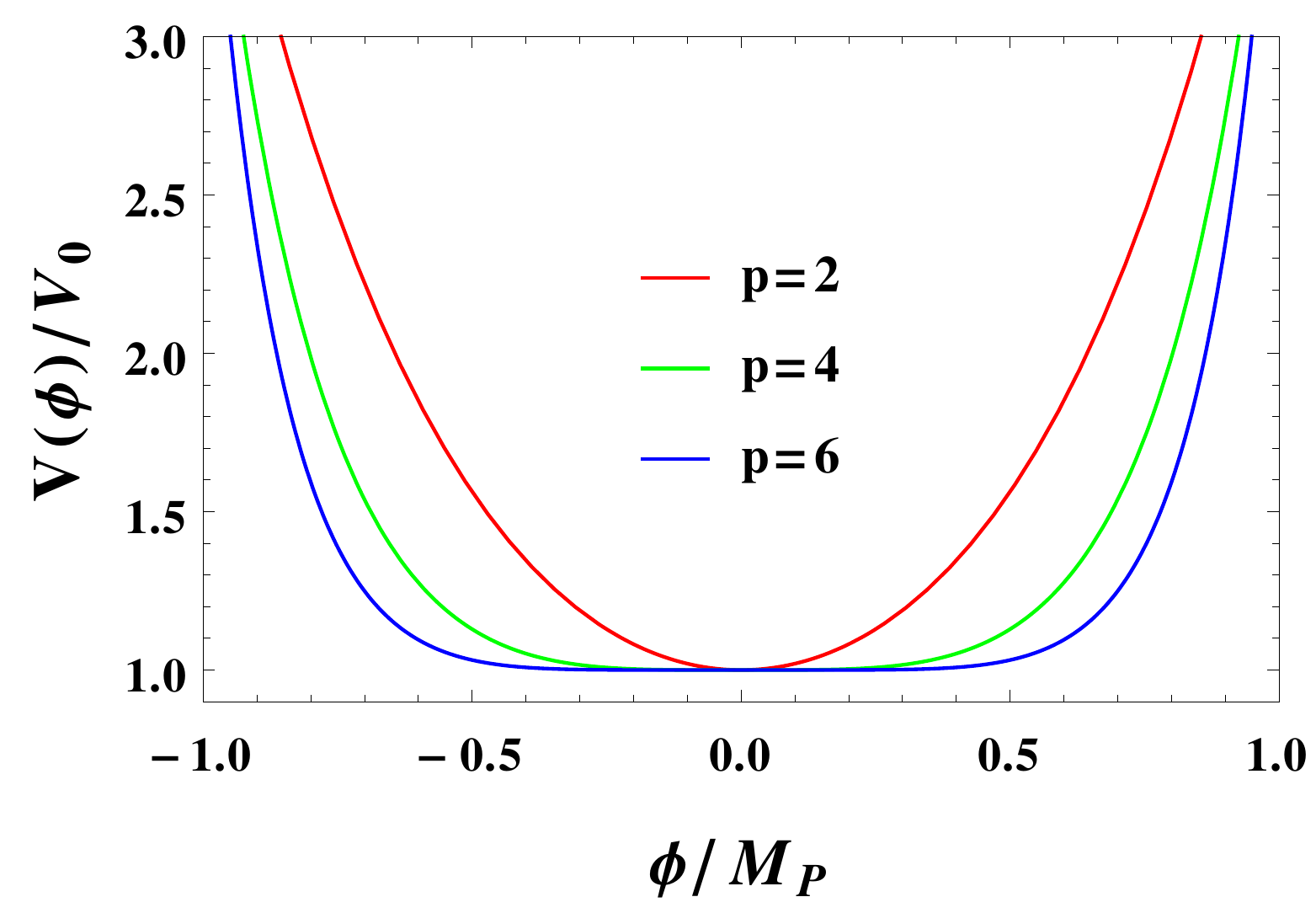}} &
{\includegraphics[width=2.5in,height=2.5in,angle=0]{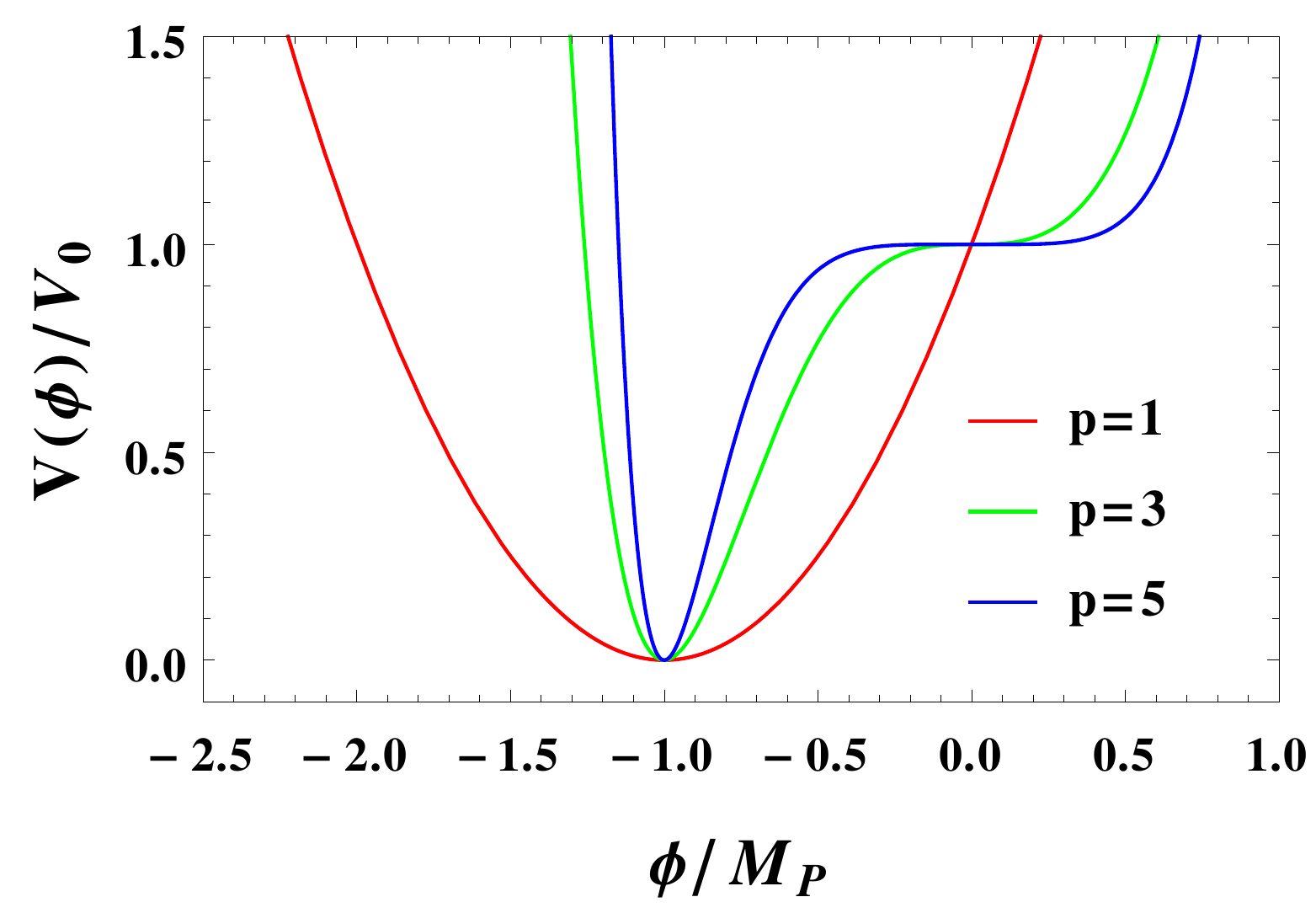}} 
\end{tabular}
\end{center}
\caption{ The figure exhibits the evolution of potential (\ref{eq:model1}) versus field $\phi$. For even (odd) values of $p$, potential shows minima at $\phi=0$ ($\phi=-1$).}
\label{fig:pot1}
\end{figure}
For model 1, we consider following potential \cite{yang}
\begin{eqnarray}
V(\phi) = V_0 (1+ \phi^p)^2
\label{eq:model1}
\end{eqnarray}

In these specific forms, $b={B''(\phi)\phi}/{B'(\phi)}=1$ and
  $c={V'(\phi)\phi}/{V(\phi)}=\frac{2p A^{p/2}}{(2+6 \xi A)^{p/2}+ A^{p/2}}$. For $b=1$, a simple relation exists between $x$ and $z$ which is given by
\begin{eqnarray}
\label{eq:b1} x&=&\frac{\dot{\phi^2}}{H^2(1-6\xi B(\phi))}
=\frac{\dot{\phi^2}}{H^2(1-6\xi B(\phi))}\frac{{(6\xi
B'(\phi))}^2}{{(6\xi B'(\phi))}^2} \frac{\phi(1-6\xi
B(\phi))}{\phi(1-6\xi B(\phi))}=\frac{z^2}{72\xi^2  A}
\end{eqnarray}

We put equation (\ref{eq:b1}) and $b=1$ in  equation (\ref{eq:capxy}), and find
 \begin{eqnarray}
\label{eq:b3}
\Omega&=&1-\frac{z^2}{72{\xi}^2 A}-y-z,\nonumber\\
X&=&-\frac{z}{2}-\frac{1}{1+18\xi^2 A}\left (\frac{z^2}{12}(6\xi -1)+y\xi (12\xi A+c) +3\xi^2 A \Omega (1-3\omega)\right ),\nonumber\\
Y&=&\frac{1}{1+18\xi^2  A}\left (\frac{z^2}{12\xi^2 A}(6\xi -1)+6y (2-3 c\xi )+3 \Omega (1-3\omega)\right).
\end{eqnarray}
Substituting equations (\ref{eq:b1}) and (\ref{eq:b3}) in (\ref{eq:autonomous4}), we finally obtain 
\begin{eqnarray}
\label{eq:b4}
y'&=&\frac{yz}{6\xi}\frac{c}{A}-2y\Big{(}\frac{1}{6(1+18\xi^2  A)}\left (\frac{z^2}{12\xi^2 A}(6\xi -1)+6y
 (2-3 c\xi ) +3(1-\frac{z^2}{72\xi^2  A}-y-z)(1-3\omega)\right )-2\Big{)}
+yz, \nonumber \\
z'&=&\left( -3z-\frac{6}{1+18\xi^2  A}\Big{(}\frac{z^2}{12}(6\xi
-1)+y\xi (12\xi A+c)+3\xi^2 A(1-\frac{z^2}{72\xi^2
A}-y-z)(1-3\omega) \Big{)}\right)\nonumber\\
&& +\frac{z^2}{6\xi A}-z\left( \frac{1}{6(1+18\xi^2
A)}\Big{(}\frac{z^2}{12\xi^2 A}(6\xi -1)+6y (2-3 c\xi )
+3(1-\frac{z^2}{72\xi^2  A}-y-z)(1-3\omega) \Big{)}-2 \right)
+z^2,\nonumber\\
 A'&=&\frac{z}{3\xi}+Az.
\end{eqnarray}
Now, the stationary points can be obtained by equating the left hand side of (\ref{eq:b4}) to zero, and the stability can be found by the sign of corresponding eigen values that will be obtained numerically. Hence, we find following stationary points for $p=2$.

\begin{enumerate}
\item 

\begin{eqnarray}
\label{eq:point1}
y&=&1, \qquad z=0, \qquad A=-\frac{1}{3 \xi}, \qquad \Omega=0,
\end{eqnarray}
The corresponding eigenvalues are given by,
\begin{eqnarray}
{\mu}_1 &=&0, \qquad
 {\mu}_2 =-3, \qquad
  {\mu}_3 = -3 (1+w),
\end{eqnarray}
One of the eigenvalue is zero. Hence, this point is not stable in the usual sense.

In this case $Y=12$ and the expression of scale factor can be obtained by using equation (\ref{eq:xyr})
 \begin{eqnarray}
Y&=&\frac{R}{H^2}=6\left(2+\frac{\dot{H}}{H^2}\right)=12
\end{eqnarray}
which tells us that 
\begin{eqnarray}
\frac{\dot{H}}{H^2}=0,
\end{eqnarray}
and finally, we have
\begin{eqnarray}
\label{eq:a}
a(t)&=&a_0 e^{H_0(t-t_0)}
\end{eqnarray}
To get the expression of $\phi(t)$, we use the following combination of the dimensionless variables
\begin{equation}
\label{Eq:beta}
\frac{z}{6\xi A}=\frac{\dot{\phi}}{\phi H}
\end{equation}
For the stationary point, $z=0$ which implies that $\dot{\phi}=0$, and hence $\phi=\phi_0$ (constant). One can see that $\dot{H}=\dot{\phi}=0$, the point shows de-Sitter behavior but looking the eigenvalues, it does not reflect a stable de-Sitter in the usual sense.

\item
\begin{eqnarray}
\label{eq:point2}
y&=&0, \qquad z=\frac{4 \xi (1-3w)}{1-w-4 \xi}, \qquad A=-\frac{1}{3 \xi}, \qquad \Omega=1- \frac{4 \xi (1-3w)}{1-w-4 \xi}+ \frac{2(3w\xi-\xi)^2}{3\xi (4\xi+w-1)^2},
\end{eqnarray}
The corresponding eigenvalues are given by,
\begin{eqnarray}
{\mu}_1 &=&\frac{4 \xi (1-3w)}{1-w-4 \xi}<0, \qquad \text{for} \qquad 4 \xi (1-3w)<0 \nonumber\\
 {\mu}_2 &=&\frac{3-16 \xi+3w(8 \xi+w-2)}{2(4 \xi+w-1)}<0, \qquad \text{for} \qquad 3+3w(8 \xi+w-2)<16 \xi \nonumber\\
  {\mu}_3 &=& \frac{-3+3w(w-4 \xi)+20 \xi}{4 \xi+w-1}0, \qquad \text{for} \qquad 3w(w-4 \xi)+20 \xi <3
\end{eqnarray}
The eigenvalues show negativity for above mentioned conditions. Therefore, this is a stable point. For this point, we have $Y=\frac{3(1-w)(1-3w)}{1-w-4 \xi}$. The time dependence of the scale factor can be obtained by using equation (\ref{eq:xyr})
 \begin{eqnarray}
Y&=&\frac{R}{H^2}=6\left(2+\frac{\dot{H}}{H^2}\right)=\frac{3(1-w)(1-3w)}{1-w-4 \xi}
\end{eqnarray}
On integrating above equation, one can easily find the expression of $a(t)$:
\begin{eqnarray}
\label{eq:apoint2a}
a(t)&=&a_0 \mid t-t_0 \mid^{\frac{1}{2-\frac{Y}{6}}}
\end{eqnarray}
where $a_0$ and $t_0$ are integration constant. For the stationary point under consideration, we finally have
\begin{eqnarray}
\label{eq:apoint2b}
a(t)&=&a_0 \mid t-t_0 \mid^{\frac{2(1-w-4 \xi)}{4(1-4 \xi)-3w^2}}
\end{eqnarray}
For the expression of $\phi(t)$, we use equation (\ref{Eq:beta}) with this stationary point, and get
\begin{eqnarray}
\label{eq:phipoint2}
\phi(t)&=& \phi_0 \mid t-t_0 \mid^{-\frac{4 \xi(1-3w)}{4(1-4 \xi)-3w^2}}
\end{eqnarray}
where $\phi_0$ and $t_0$ are integration constant. The expressions of $a(t)$ and $\phi(t)$ give power law solutions that do not satisfy the de-Sitter condition. 

\item
\begin{eqnarray}
\label{eq:point3}
y&=&0, \qquad z=12 \xi-2 \sqrt{36 \xi^2-6 \xi}, \qquad A=-\frac{1}{3 \xi}, \nonumber\\  \Omega &=& 1- 12 \xi+ 2  \sqrt{36 \xi^2-6 \xi}+ \frac{(6\xi- \sqrt{36 \xi^2-6 \xi})^2}{6\xi },
\end{eqnarray}
The corresponding eigenvalues are following, and show negativity for below conditions.
\begin{eqnarray}
{\mu}_1 &=& 12 \xi- 2  \sqrt{36 \xi^2-6 \xi}<0, \qquad \text{for} \qquad 2  \sqrt{36 \xi^2-6 \xi}<12 \xi \nonumber\\
 {\mu}_2 &=& 3-3w- 12 \xi+ 2  \sqrt{36 \xi^2-6 \xi}<0, \qquad \text{for} \qquad 3+2  \sqrt{36 \xi^2-6 \xi}<3w + 12 \xi \nonumber\\
  {\mu}_3 &=& 6-36 \xi+6   \sqrt{36 \xi^2-6 \xi}<0, \qquad \text{for} \qquad 6+6  \sqrt{36 \xi^2-6 \xi}<36 \xi \nonumber\\
\end{eqnarray}
In this case $Y=-6+72 \xi- 12  \sqrt{36 \xi^2-6 \xi}$, and the expression of $a(t)$ and $\phi(t)$ are found to be
\begin{eqnarray}
\label{eq:apoint3}
a(t)&=&a_0 \mid t-t_0 \mid^{\frac{1}{3-12\xi+2\sqrt{36 \xi^2-6 \xi}}}
\end{eqnarray}
For the expression of $\phi(t)$, we use equation (\ref{Eq:beta}) with this stationary point, and get
\begin{eqnarray}
\label{eq:phipoint3}
\phi(t)&=& \phi_0 \mid t-t_0 \mid^{\frac{-6 \xi+\sqrt{36 \xi^2-6 \xi}}{3-12\xi+2\sqrt{36 \xi^2-6 \xi}}}
\end{eqnarray}
Again expressions $a(t)$ and $\phi(t)$ provide power law solutions that do not qualify for de-Sitter solution.

\end{enumerate}

\subsubsection{Model 2: $B(\phi) \propto {\phi}^2$, $V(\phi) = V_0 e^{\lambda \phi^2}$}
\label{sec:model2}
For model 2, we consider following potential \cite{alamEPJC}
\begin{eqnarray}
V(\phi) =  V_0 e^{\lambda \phi^2}
\label{eq:model2}
\end{eqnarray}
Equations (\ref{eq:b3}) and (\ref{eq:b4}) will remain same except the dimensionless parameter $c$. For model 2, the expression of $c$ is given as
\begin{eqnarray}
c = \frac{\lambda A}{1+3 \xi A}
\label{eq:c}
\end{eqnarray}

Now, the stationary point can be obtained by equating the left hand side of (\ref{eq:b4}) to zero, and the stability can be found by the sign of corresponding eigen values that will be obtained numerically. Hence, we find following stationary point.
\begin{eqnarray}
\label{eq:point}
y&=&1, \qquad z=0, \qquad A=-\frac{\lambda + 12\xi}{36 \xi^2}, \qquad \Omega=0,
\end{eqnarray}
The corresponding eigenvalues are given by,
\begin{eqnarray}
{\mu}_1 &=&-3(1+w),\\
 {\mu}_2 &=&-\frac{3}{2}-\frac{1}{2}\sqrt{\frac{18+7 \alpha+84 \xi}{2-\alpha-12 \xi}},\\
  {\mu}_3 &=&-\frac{3}{2}+\frac{1}{2}\sqrt{\frac{18+7 \alpha+84 \xi}{2-\alpha-12 \xi}},\\
\end{eqnarray}
Above eigenvalues show negativity for the below conditions.
\begin{eqnarray}
{\mu}_1 & < & 0 ~\text{for}~ w> -1\\
 {\mu}_2 & < & 0 ~\text{for}~ \sqrt{\frac{18+7 \alpha+84 \xi}{2-\alpha-12 \xi}}> 0\\
  {\mu}_3 & < & 0 ~\text{for}~  \sqrt{\frac{18+7 \alpha+84 \xi}{2-\alpha-12 \xi}} < 0.
\end{eqnarray}

In this case $Y=12$ and the expression of scale factor can be obtained by using equation (\ref{eq:xyr})
 \begin{eqnarray}
Y&=&\frac{R}{H^2}=6\left(2+\frac{\dot{H}}{H^2}\right)=12\nonumber
\end{eqnarray}
which tells us that 
\begin{eqnarray}
\frac{\dot{H}}{H^2}=0,
\end{eqnarray}
and finally, we have
\begin{eqnarray}
\label{eq:a}
a(t)&=&a_0 e^{H_0(t-t_0)}
\end{eqnarray}
To get the expression of $\phi(t)$, we use the following combination of the dimensionless variables
\begin{equation}
\frac{z}{6\xi A}=\frac{\dot{\phi}}{\phi H}
\end{equation}
For the stationary point, $z=0$ which implies that $\dot{\phi}=0$, and hence $\phi=\phi_0$.

where $a_0$, $t_0$ are integration constants, and  $\phi_0=\pm\sqrt{\frac{\lambda + 12\xi}{6\xi \lambda}}$,
$H_0=\pm\sqrt{-\frac{\lambda e^{\lambda \phi_0^2}}{6\xi }}$ are obtained
from the system (\ref{eq:Friedphi})- (\ref{eq:KGphi}) for
$\dot{H_0}=\dot{\phi_0}=\ddot{\phi_0}=\rho=0$. 

we next consider the behavior of $G_{eff}$, and can be written as in terms of dimensionless parameters
\begin{equation}
G_{eff}=\frac{\kappa}{8 \pi (1-\kappa \xi B(\phi))}=\frac{\kappa \alpha A}{8 \pi c}.
\end{equation}

\begin{figure}
  \begin{center}
     \includegraphics[width=3in,height=3in,angle=0]{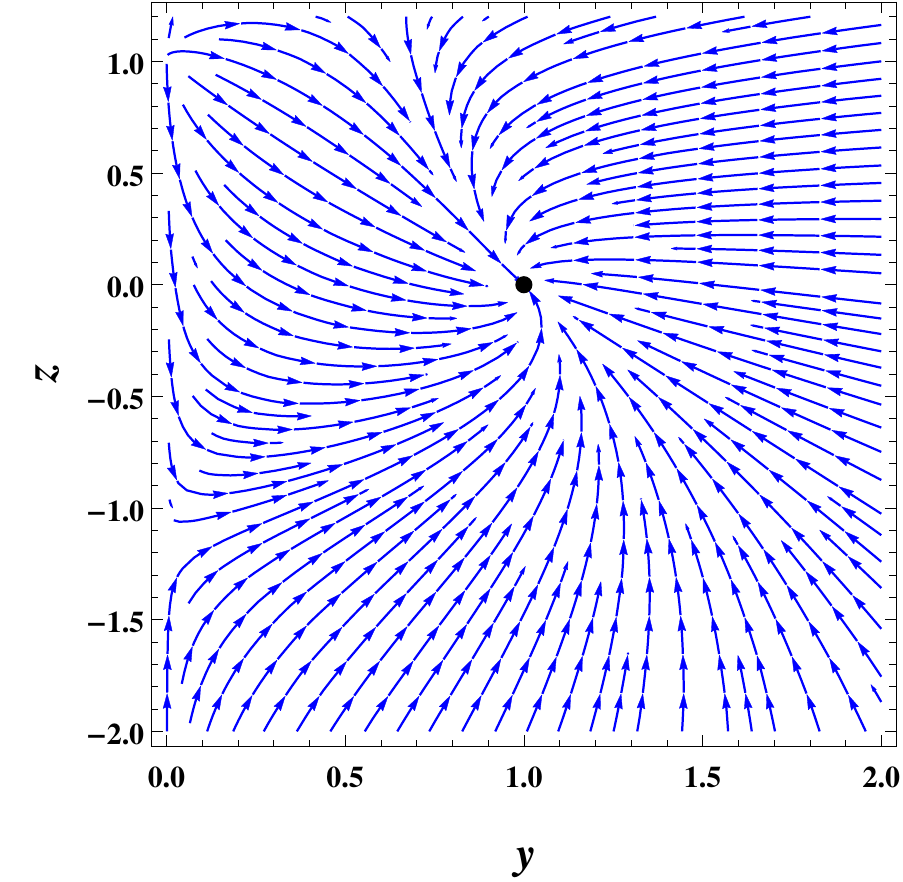}
     \end{center}
\caption{\small The figure represents the stable fixed point for model 2 with $\lambda=2$, $\xi=2$ and $w=0$. The eigen values corresponding to the chosen parameters are $\mu_1=-3$, $\mu_2=-1.5 - 1.44338 i$ and $\mu_3=-1.5 + 1.44338 i$ which exhibit that the stable point is an attractive focus. The black dot shows the stable attractor point.}
\label{fig:port}
\end{figure}

\begin{figure}[tbp]
\begin{center}
\begin{tabular}{ccc}
{\includegraphics[width=2.5in,height=2.5in,angle=0]{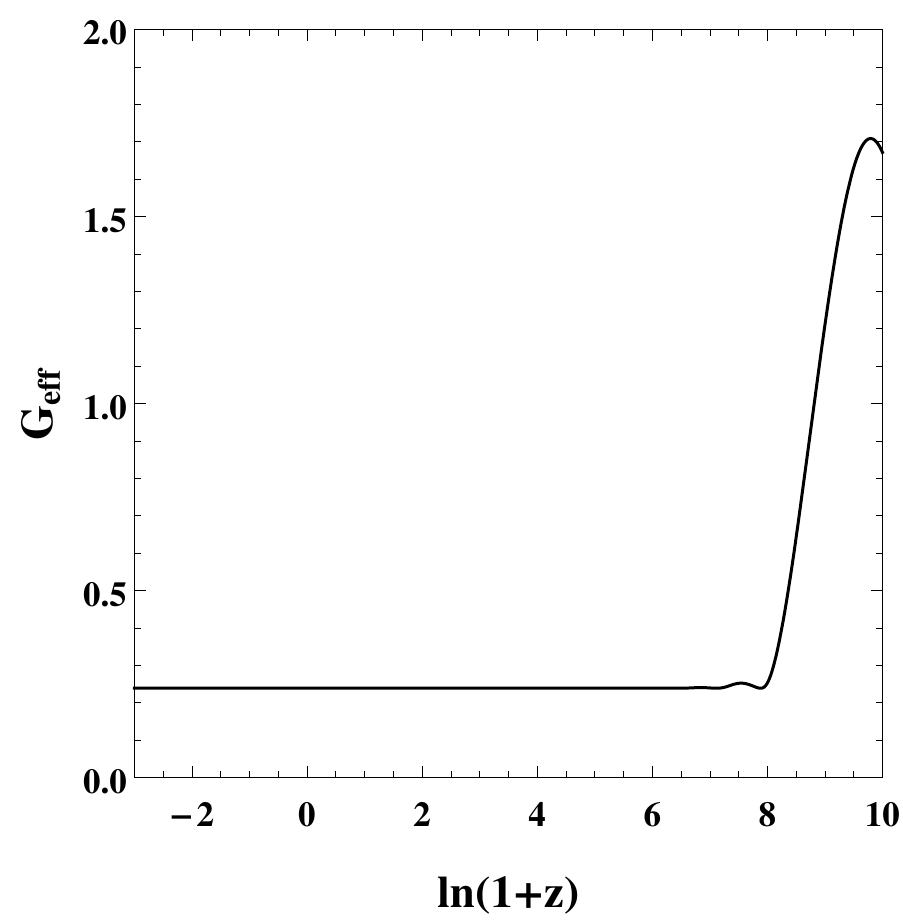}} &
{\includegraphics[width=2.5in,height=2.5in,angle=0]{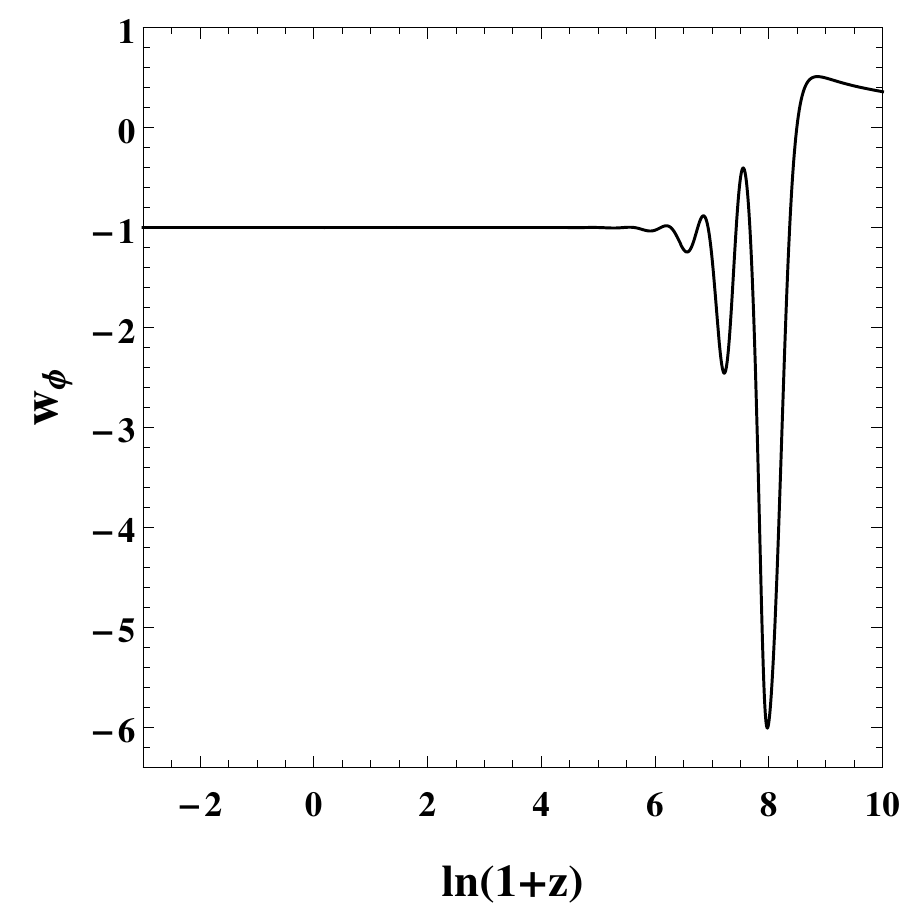}} 
\end{tabular}
\end{center}
\caption{ The figure shows the evolution of $G_{eff}$ and $w_{\phi}$ versus redshift $z$ for model 2. The $G_{eff}$ remains positive during the whole evolution whereas $w_{\phi}$ passes through a brief phantom phase in the past and reaches at $w_{\phi} \simeq -1$ around the present epoch that provides a stable de-Sitter solution. We choose $\lambda=2$, $\xi=2$ and $w=0$ in the numerical evolution.}
\label{fig:GN}
\end{figure}

It is remarkable to see that in the absence of curvature term $B(\phi)R $ in the action, a family of de-Sitter solutions are obtained \cite{topo}. Our numerical work shows that in the presence of curvature term $B(\phi)R $, we also obtain a true de-Sitter solution for $B(\phi) \propto \phi^2$ and $V(\phi) \propto e^{\lambda \phi^2}$  as $G_{eff}$ and $\phi$ are constant. 

A de-Sitter solution corresponds to $\dot{H}=0$, $\phi=$ constant, $w_{\phi} \simeq -1$ and $G_{eff}=$ constant which is in our case trivially satisfied. However, in our previous study, a true de-Sitter solution is not captured by the current autonomous system for $B(\phi) \propto \phi^N$ and $V(\phi) \propto  \phi^n$ \cite{alam2012}.

The phase space trajectories for the stable point [Eq. (\ref{eq:point})] for $\lambda=2$, $\xi=2$ and $w=0$ are exhibited in Fig. \ref{fig:port} for which the eigen values are $\mu_1=-3$, $\mu_2=-1.5 - 1.44338 i$ and $\mu_3=-1.5 + 1.44338 i$. In this case, all trajectories move towards the stable attractor point, and it behaves as an  attractive focus. 

The evolution of $G_{eff}$ and $w_{\phi}$ versus redshift $z$ are shown in Fig. \ref{fig:GN} in which the evolution of $G_{eff}$ remains positive throughout the evolution. Before approaching the stable de-Sitter, the equation of state pass through a brief phantom phase in the past, and at present epoch it gives observed value $w_{\phi} \simeq -1$ with $G_{eff}>0$.

\section{Conclusion}
\label{sec:conc}
In this paper, we have re-performed a dynamical analysis for a NMC scalar field model described by $F(\phi)R= (1-\xi B(\phi))R $ with $B(\phi) \propto \phi^2$ and $V(\phi) = V_0 (1+ \phi^p)^2$, $V(\phi) =V_0 e^{\lambda \phi^2}$ using a suitable set of dimensionless variables. We used an autonomous system for a scalar-tensor model of dark energy with nonminimial coupling that account for late time cosmic acceleration. This work is similar to our previous paper \cite{alam2012}. In our earlier work, we obtained a transient phase of dark energy and a de-Sitter solution with $G_{eff}<0$, and if a universe with $G_{eff}<0$ exists then it will be different from our real universe. In the current study, we use same autonomous system with $B(\phi) \propto \phi^2$ and power law, steep exponential potentials, a true de-sitter solution with $G_{eff}>0$ is found as $\dot{H}=0$, $w_{\phi} \simeq -1$ and $\phi=$ constant. However, in case of power law potential, one of the stationary point satisfies the de-Sitter conditions but the point is not stable in actual sense.

The phase space trajectories of a stable point are presented in Fig. \ref{fig:port}. All trajectories around the current epoch converges to $w_{\phi} \simeq -1$, and reaches to the stable attractor point that behaves as an attractive focus. In the left panel of Fig. \ref{fig:GN}, we have shown the evolution of $G_{eff} $ versus redshift that gives $G_{eff}>0$ during the entire evolution. In the right panel of Fig. \ref{fig:GN}, we exhibited the evolution of $w_{\phi}$ that provides the transient phantom phase in the past and de-Sitter solution around the current epoch before moving towards the stable attractor point.

\section*{Acknowledgments}
The work by MK was supported by the Ministry of Education and Science of Russian Federation, MEPhI Academic Excellence Project (contract 02.a03.21.0005, 27.08.2013).


\begin{thebibliography}{99}

\bibitem{review2}E.~J.~Copeland, M.~Sami and
S.~Tsujikawa, Int. J. Mod. Phys., {D15} ,
1753(2006)[hep-th/0603057].

\bibitem{alamLQC} M. Shahalam, M. Sami, A. Wang, Phys. Rev. D 98,  043524 (2018);
M. Sharma, M. Shahalam, Q. Wu, A. Wang, JCAP 11 (2018) 003; M. Shahalam, M. Sharma, Q. Wu, A. Wang, Phys. Rev. D 96, 123533 (2017);  M. Shahalam, Universe 4 (2018) 87.

\bibitem{wangLQC} A. García-Quismondo, G. A. Mena Marugán,  	Phys. Rev. D 99, 083505 (2019) [arXiv:1903.00265]; Bao-Fei Li, P. Singh, A. Wang [arXiv:1906.01001].


\bibitem{BD} C. Brans and R. Dicke, Phys. Rev.  {\bf 124}, 925 (1961).
\bibitem{PR1} S. Perlmutter {\it el al}, Astrophysics, J. {\bf 157},
565(1999).
\bibitem{PR2} A. Reiss {\it et al}, Astrophysics J. {\bf 117},
707(1999).

\bibitem{review1}
V.~Sahni and A.~A.~Starobinsky, Int.\ J.\ Mod.\ Phys.\ D \textbf{9},
373 (2000).
\bibitem{vpaddy} V. Sahni and A. Starobinsky,
     Int.J.Mod.Phys.D {\bf 15}, 2105(2006)[astro-ph/0610026]; T. Padmanabhan,
     astro-ph/0603114;
 P.~J.~E.~Peebles and B.~Ratra, Rev.\ Mod.\ Phys.\ {} \textbf{75},
559 (2003); L. Perivolaropoulos, astro-ph/0601014; N. Straumann,
arXiv:gr-qc/0311083;  J. Frieman, arXiv:0904.1832; M. Sami, Lect.
Notes Phys.{\bf 72}, 219(2007); M. Sami,  arXiv:0901.0756 ; K.
Bamba, S. Capozziello, S. Nojiri and S. D. Odintsov,
arXiv:1205.3421; S. Tsujikawa, arXiv:1004.1493.


\bibitem{review3} E. V. Linder,  Rep. Prog. Phys. {\bf 71} (2008)
056901.
\bibitem{review3C} Robert R. Caldwell and Marc
Kamionkowski,arXiv:0903.0866.
\bibitem{review3d} A. Silvestri and Mark Trodden, arXiv:0904.0024.
\bibitem{review4}J. Frieman, M. Turner and D.
Huterer, arXiv:0803.0982.
\bibitem{a1} I. Ya. Aref'eva, N. V. Bulatov , R. V. Gorbachev, S. Yu. Vernov,  Class. Quant. Grav. 31 (2014) 065007 [arXiv:1206.2801].

\bibitem{a2} A. Yu. Kamenshchik, A. Tronconi, G. Venturi, S. Yu. Vernov,  Phys.Rev. D87 (2013) no.6, 063503 [arXiv:1211.6272].

\bibitem{a3} K. Nozari, N. Rashidi, Astrophys.Space Sci. 347 (2013) 375-388  [arXiv:1308.5772].

\bibitem{a4} J. B. Dent, S. Dutta, E. N. Saridakis, Jun-Qing Xia, JCAP 1311 (2013) 058  [arXiv:1309.4746]. 


\bibitem{a5} Alexander Yu. Kamenshchik, Ekaterina O. Pozdeeva, Alessandro Tronconi, Giovanni Venturi, Sergey Yu. Vernov,  Class.Quant.Grav. 31 (2014) 105003  [arXiv:1312.3540].

\bibitem{a6} Xiangzhong Luo, Puxun Wu, Hongwei Yu.,  Astrophys.Space Sci. 350 (2014) no.2, 831-837.

\bibitem{a7} E. Elizalde, S.D. Odintsov, E.O. Pozdeeva, S. Yu. Vernov, Phys.Rev. D90 (2014) no.8, 084001  [arXiv:1408.1285].

\bibitem{a8} Yumei Huang, Qing Gao, Yungui Gong,  Eur.Phys.J. C75 (2015) no.4, 143  [arXiv:1412.8152]; Yumei Huang, Yungui Gong, Dicong Liang, Zhu Yi, Eur.Phys.J. C75 (2015) no.7, 351  [arXiv:1504.01271]; Nan Yang, Qin Fei , Qing Gao, Yungui Gong, Class.Quant.Grav. 33 (2016) no.20, 205001 [arXiv:1504.05839]; Nan Yang, Qing Gao, Yungui Gong, Int.J.Mod.Phys. A30 (2015) no.28n29, 1545004; Yi Zhu, Yungui Gong,  Int.J.Mod.Phys. D26 (2016) no.02, 1750005  [arXiv:1512.05555]  

\bibitem{a9} Sebastian Bahamonde, Matthew Wright,  Phys.Rev. D92 (2015) no.8, 084034 [arXiv:1508.06580].

\bibitem{a10} A. Yu. Kamenshchik, E.O. Pozdeeva, A. Tronconi, G. Venturi, S. Yu. Vernov, Class.Quant.Grav. 33 (2016) no.1, 015004  [arXiv:1509.00590]. 


\bibitem{a11} Somnath Bhattacharya, Pradip Mukherjee, Amit Singha Roy, Anirban Saha,  Eur.Phys.J. C78 (2018) no.3, 201  [arXiv:1512.03902].

\bibitem{a12} Behnaz Fazlpour, Gen.Rel.Grav. 48 (2016) no.12, 159  [arXiv:1604.03080].

\bibitem{a13} Tiberiu Harko, Francisco S. N. Lobo, Emmanuel N. Saridakis, Minas Tsoukalas,   Phys.Rev. D95 (2017) no.4, 044019  [arXiv:1609.01503].

\bibitem{a14} Pradip Mukherjee, Anirban Saha, Amit Singha Roy,  Mod.Phys.Lett. A33 (2018) no.02, 1850010  [arXiv:1609.04752].

\bibitem{a15} Arvin Ravanpak, Hossein Farajollahi, Golnaz Fadakar, Res.Astron.Astrophys. 16 (2016) no.9, 137 [arXiv:1610.09614].


\bibitem{a16} L.N. Granda, D.F. Jimenez, Eur.Phys.J. C77 (2017) no.10, 679  [arXiv:1710.04760]; Int.J.Mod.Phys. D27 (2017) no.03, 1850030  [arXiv:1710.07273].

\bibitem{sunny} R. Myrzakulov, L. Sebastiani, S. Vagnozzi, Eur. Phys. J. C75 (2015) 444.

\bibitem{Polarski} R.Gannouji, D.Polarski, A.Ranquet and A.Starobinsky, JCAP 0609:016 (2006).

\bibitem{alam2012} M. Sami, M. Shahalam, M. Skugoreva, A. Toporensky, Phys.Rev. {\bf D86},  103532 (2012).


\bibitem{yang} W. Yang, M. Shahalam, B. Pal, S. Pan, A. Wang [{arXiv:1810.08586}].

\bibitem{alamEPJC} M. Shahalam, W. Yang, R. Myrzakulov, A. Wang,  Eur. Phys. J. C77 (2017) no.12, 894 [arXiv:1802.00326].

\bibitem{topo} A. Yu. Kamenshchik, I. M. Khalatnikov, and A. V. Toporensky, Int. J. Mod. Phys. D 06, 649 (1997).

\end{thebibliography}
\end{document}